\documentclass[11pt]{article}
\usepackage{amsmath,amsfonts,amssymb,latexsym}

\newtheorem{theorem}{Theorem}[section]

\numberwithin{equation}{section}
\def\be{\begin{equation}}
\def\ee{\end{equation}}
\def\bq{\begin{eqnarray}}
\def\eq{\end{eqnarray}}
\def\beq{\begin{eqnarray*}}
\def\eeq{\end{eqnarray*}}

\def\r{\rho}
\def\a{\alpha}
\def\b{\beta}
\def\g{\gamma}
\def\G{\Gamma}
\def\d{\delta}
\def\l{\lambda}
\def\m{\mu}

\def\z{\zeta}

\newcommand{\GA}{\alpha}
\newcommand{\GB}{\beta}
\newcommand{\GG}{\gamma}

\begin{document}
\title{\huge{Structure of infinity in cosmology}}
\author{\Large{ Spiros Cotsakis\footnote{\texttt{email:\thinspace skot@aegean.gr}} } \\
{\normalsize Research Group of Geometry, Dynamical Systems and
Cosmology}\\ {\normalsize Department of Information and
Communication Systems Engineering}\\ {\normalsize University
of the Aegean}\\ {\normalsize Karlovassi 83 200, Samos,
Greece}}
\maketitle
\begin{abstract}
\noindent We discuss recent developments related to certain blow up methods suitable for the analysis of cosmological singularities and asymptotics. We review results obtained in a variety of currently popular themes and  describe ongoing research about universes with various kinds of extreme states, higher order gravity, and certain models of braneworlds.
\end{abstract}

\tableofcontents

\section{Introduction}	
The phrase `infinity in cosmology' included in the title of this paper may mean completely different and mutually exclusive things. For example: `Will this or that property remain valid forever?', or `why some solution disappeared after a very short period of existence?', are questions of great  interest for the mathematical cosmologist. The purpose of this paper is to present  methods and results of recent and ongoing research that may help to address these and other similar questions in a cosmological setting.

The plan of this paper is as follows. In Section 2, we present an overview of basic approaches to the problem of the description of singularities and asymptotics for cosmological spacetimes. The next three sections, provide more than just applications of these mathematical techniques to a variety of problems in theoretical and mathematical cosmology. In Section 3, we consider the behaviour of universes, coined here `extremophilic' (an adaptation from mathematical biology),  having an `extreme state' and develop  the relevant theory for collapse singularities, sudden ones,  dark energy universes with big rips,  and also those that depend on an interacting pair of fluids. Section 4 focuses on universes in higher order gravity, in particular, those in vacuum and those filled with radiation. We also discuss the question of genericity of regular states in these theories. Finally, in Section 5, we review  recent results in brane cosmology obtained under specific assumptions about the bulk content, with particular emphasis in the relation to the self-tuning mechanism.

In this paper we use the following notation. We consider a spacetime $(\mathcal{V},g)$ where $\mathcal{V}=\mathbb{R}\times\mathcal{M}$, with $\mathcal{M}$ being an orientable 3-manifold, the submanifolds
$\mathcal{M}_{t}=\{t\}\times \mathcal{M}$, $t\in\mathbb{R}$, are spacelike and $g$ is a Lorentzian metric, analytic and
with signature $(+,-,-,-)$. We take a \emph{Cauchy adapted frame}, cf. Ref. \cite{ycb}, $e_i=(e_{0},e_{\a})$
with $e_{\a}$ tangent to the space slice $\mathcal{M}_{t}$ and
$e_{0}$ orthogonal to it. The \emph{dual coframe}
$\theta^{i}=(\theta^{0}=dt,\theta^{\a}=dx^{\a}+\beta^{\a}dt)$,
 where the tangent vector
$\beta^{\a}$ is the usual \emph{shift}, leads to the standard general form
of the metric $g$,
\begin{equation}\label{metric}
  ds^2=N^{2}dt^{2}-\g_{\a\b}(t)\left(
  dx^{\a}+\beta^{\a}dt\right)\left( dx^{\b}+\beta^{\b}dt\right) .
\end{equation}
Here $N$ is a positive function,  the \emph{lapse}, and we assume
that all metrics $\g_{\a\b}(t)$ are complete Riemannian metrics with $\g_{\a\b}=-g_{\a\b}$ and
$\g^{\a\b}=-g^{\a\b}$.  Below we shall deal exclusively with the simplest gauge choice, $N=1, \b=0$, which means that $ g_{00}=1, g_{0\GA}=0$, and local coordinates are \emph{adapted} to the product structure on $\mathcal{M}_{t}$, $(x^i=(t,x^\a))$ (in this case, one sometimes speaks of a \emph{synchronous} system of local coordinates). We define the extrinsic curvature of $(\mathcal{M},\g)$ through the \emph{first variational equation}
\be\label{eq:pg}
\partial_t \g_{\GA\GB} = K_{\GA\GB}.
\ee
We shall also be interested in Section 4 in the dynamics of FRW universes near various  singularities and asymptotic regimes. These cosmologies are determined by the Robertson-Walker metric of the form
\be \label{rwmetrics}
g_{4}=dt^{2}-a^{2}\, g_{3},
\ee
where $a(t)$ denotes the scale factor, while each slice is given the 3-metric
$
g_{3}=(1-kr^2)^{-1}dr^{2}+r^2g_{2},
$
$k$ being the (constant) curvature normalized to take the three values $0, +1$ or $-1$ for the complete, simply connected,  flat, closed or open space sections respectively, with the 2-dimensional sections having the metric $
g_{2}=d\theta^{2}+\sin^{2}\theta d\phi^{2}.$

Lastly, another type of model cosmology  will appear in Section 5. We shall require the basic geometry of a braneworld model consisting of a three-brane embedded
in a five-dimensional bulk space that is filled with a massless
scalar field and/or an analog of perfect fluid. We shall assume a bulk metric of the
form \be \label{warpmetric} g_{5}=a^{2}(Y)g_{4}+dY^{2}, \ee
where
$g_{4}$ is the four-dimensional flat, de Sitter or anti de Sitter
metric, i.e., $g_{4}=-dt^{2}+f^{2}_{\kappa}g_{3}, $ where $g_{3}=dr^{2}+h^{2}_{\kappa}g_{2} $
and $g_{2}=d\theta^{2}+\sin^{2}\theta d\varphi^{2}.$
Here $f_{\kappa}=1,\cosh (H t)/H,\cos (H t)/H$
($H^{-1}$ is the de Sitter curvature radius) and $h_{\kappa}=r,\sin r,\sinh r$,
respectively. $Y$ denotes the fifth dimension.
For the scalar field we shall assume an energy-momentum tensor of
the form $T^{1}_{AB}=(\rho_{1}+P_{1})u_{A}u_{B}-P_{1}g_{AB}$,
where $A,B=1,2,3,4,5$, $u_{A}=(0,0,0,0,1)$, and $\rho_{1}$, $P_{1}$
are the density and pressure of the scalar field,
$P_{1}=\rho_{1}=\lambda\phi'^{2}/2$. Here, the prime
denotes differentiation with respect to $Y$, and $\lambda$ is a
parameter. The energy-momentum tensor of the fluid
is $T^{2}_{AB}=(\rho_{2}+P_{2})u_{A}u_{B}-P_{2}g_{AB}$, and we assume an
equation of state of the form $P_{2}=\gamma\rho_{2}$ between the
pressure $P_{2}$ and the density $\rho_{2}$ with $\gamma$ being a parameter.
All quantities $\rho_{1}$, $\rho_{2}$ and $P_{1}$, $P_{2}$ will be functions of the fifth
dimension $Y$ only.

\section{Blow up  methods}
In this section, we present basic ideas behind four different methods that have proven useful for the description of cosmological asymptotics and singularities. Details of the techniques involved in all these approaches are given in the references quoted in the subsections of this section. Although they are logically independent from each other and come from completely distinct motivations, each one of these four blow up methods highlights a different aspect of the general problem, and perhaps one needs to combine different results in order to built a more complete and coherent picture.

\subsection{Classification criteria}
The  theorems quoted in this subsection were proved in Refs.  \cite{ycb-cot02,cot04,ck05,ck07,cot07}, where the reader is invited to look for the definitions of the various technical terms and further details and proofs. They are here formulated as providing sufficient conditions for a spacetime to be geodesically complete, however, we are really interested in their \emph{contrapositive} formulations, giving necessary conditions for singularity formation in the sense of geodesic incompleteness. The first theorem (proved in Ref. \cite{cot04} in the general case) eventually concerns the behaviour of the scale factor.
\begin{theorem}
Let  $(\mathcal{V},g)$ be a regularly  sliced spacetime. Then the
following are equivalent:
\begin{enumerate}
\item $(\mathcal{M}_{0},\g_{0} )$ is a $\g_0$-complete Riemannian manifold
\item The spacetime $(\mathcal{V},g)$ is globally hyperbolic
\end{enumerate}
\end{theorem}
In the case of an FRW universe, the regular slicing condition simply means that the scale factor $a$ is a bounded function of the form $0<a(t)<\infty$, and so in the contrapositive formulation this will allow two kinds of singularities, collapse singularities corresponding to  $a\rightarrow 0$, and big rip singularities, those with $a\rightarrow\infty$. We note that this theorem (cf. Ref. \cite{cot04}) implies that slice completeness is not equivalent to spacetime geodesic completeness except when the spacetime is static. Hence we have to allow for a possible third type of finite time singularity in the contrapositive formulation, that is when the scale factor takes a finite value at the time of singularity formation, $a(t_s)=a_s$. This will include  sudden singularities. We conclude that our first theorem implies the existence of three main types of possible spacetime singularities formed during the evolution as ${t\rightarrow t_{s}}$, a finite time, namely,  collapse ($a\rightarrow 0$), sudden ($a(t)\rightarrow a_s$), and big rip ($a\rightarrow\infty$) ones, denoted here with the letters $\textsc{SF}_1,\textsc{SF}_2,\textsc{SF}_3$ respectively.

It is instructive to further refine the above broad classification by using the Bel-Robinson energy (cf. Ref. \cite{ycb}), eventually defined to be  $\mathcal{B}(t)\sim |D|^{2}+ |E|^{2}$, with $|D|$ and $|E|$ being the norms of the electric parts of the Riemann double two form\footnote{the corresponding magnetic tensors are identically zero for the isotropic geometry we are considering here.}, $|D|^{2}=3\left(\left(\dot{a}/{a}\right)^{2}+k/{a^{2}}\right)^{2}$ and $|E|^{2}=3\left(\ddot{a}/{a}\right)^{2}$. We see that $D$ and $E$ are essentially proportional to the fluid density $\rho$ and pressure $p$ respectively. We then have the following result, cf. Ref. \cite{cot07}.
\begin{theorem}\label{br}
A spatially closed, expanding at time {$t_*$}, Friedmann universe
that satisfies {$d<\mathcal{B}(t)<e$}, where $d,e$ are constants,  is causally geodesically complete. Further,
there is a minimum radius, $a_{\textrm{min}}>\Delta^{-1/2}$,
$\Delta$ depending on $e$ and these universes are eternally
accelerating {($\ddot{a}>0$)}.
\end{theorem}
We therefore conclude that for each one of the three singularity types $\textsc{SF}_1,\textsc{SF}_2,\textsc{SF}_3$, Theorem \ref{br} in its contrapositive formulation introduces four further subtypes, namely,
\begin{description}
\item $\textsc{BR}_{1}:$\, $|D|< \infty \,|E|\rightarrow\infty $
\item $\textsc{BR}_{2}:$\, $|D|\rightarrow \infty,\,|E|\rightarrow\infty$
\item $\textsc{BR}_{3}:$\, $|D|< \infty ,\, |E|<\infty$
\item $\textsc{BR}_{4}:$\, $|E|<\infty,\, |D|\rightarrow \infty $
\end{description}
Of the 12 different singularity types  formed by combining the two theorems quoted above, some types have been studied in the literature in particular cases, while others are new. Note that we use the term `sudden' in a broader meaning than that introduced in Ref. \cite{ba04} (which corresponds to the type $(\textsc{SF}_2,\textsc{BR}_{1})$ above), to cover \emph{any} $\textsc{SF}_{2}$ singularity. Some properties of the three types $(\textsc{SF}_2,\textsc{BR}_{i}), i=1,2,3,$ of sudden singularities were studied in Ref.
\cite{noj1}. We are not aware  of any example of a $\textsc{BR}_{4}$ singularity.

 Any universe will belong to a particular singularity type of the 12 introduced above, according to the behaviour of its volume (as determined by the scale factor - $\textsc{SF}$ types\footnote{Note that because of the homogeneity of space, for an open or flat model, it suffices to consider any finite part of the initial spatial slice.}) and its matter content (as determined by the Bel-Robinson energy - $\textsc{BR}$ types).

 There is an extra piece of information that we may use to further refine our singularity classification, and highlight the fact that each singularity type of the form $(\textsc{SF},\textsc{BR})$ has extra structure. This is determined by the way each slice behaves on approach to the spacetime singularity, that is by the extrinsic curvature-the Hubble parameter in the case of the FRW geometry. The following completeness theorem provides the necessary background. It was first proved in Ref. \cite{ycb-cot02}.
\begin{theorem}
If $(\mathcal{V},g)$ is a globally hyperbolic, regularly sliced spacetime such that
for each finite {$t_{1},$ $|\nabla N|_{\mathbf{g}_t}$} and
{$|K|_{\mathbf{g}_t}$}  are integrable functions  on
{$[t_{1},+\infty )$}, then $(\mathcal{V},g)$ is future causally geodesically complete.
\end{theorem}
When used in each contrapositive form, this result leads to three more types of singularities based on the asymptotic behaviour of the Hubble parameter. We phrase it in future terms as in Ref. \cite{ck05}:
\begin{theorem}\label{2}
Necessary conditions for the existence of future singularities in
globally hyperbolic FRW universes are:
\begin{description}
\item $\textsc{H}_1$:\,  H is not piecewise continuous, or
\item $\textsc{H}_2$:\, H blows up in a finite time, or
\item $\textsc{H}_3$:\, H is defined and integrable  for only a finite
proper time interval.
\end{description}
\end{theorem}
We thus arrive at a complete, three-level classification of cosmological singularities based on 36 different types of the form $(\textsc{SF},\textsc{BR}, \textsc{H})$ as determined by the asymptotic  behaviours of the scale factor (volume), Bel-Robinson energy (matter fields) and Hubble parameter (extrinsic curvature) on approach to the finite time singularity. This is an asymptotic classification of cosmological singularities that takes into account not only the geometric but also the dynamical features of the fields.

\subsection{Asymptotic splittings}
In order to become more intimately acquainted with the structure of the possible singularities classified above, we focus now on a local method for the characterization of the asymptotic properties of solutions to the field
equations of a given theory of gravity in the neighborhood of the spacetime
singularity. This is the method of asymptotic splittings, cf. Ref. \cite{split}. Assume that we are given a vector field that defines the dynamical system we are about to explore, and we know that at
some point, $t_{s}$, a system of integral curves, corresponding to a
particular or a general solution, has a (future or past) finite-time
singularity.  The vector field (or its integral
curves) can basically do two things sufficiently close to the
singularity, namely, it can either show some dominant feature or not. In the
latter case, the integral curves can `spiral' in some way around the
singularity \emph{ad infinitum}, whereas in the former case solutions share a
distinctly dominant behaviour on approach to the singularity at $t_{s}$.

To describe both cases invariably, we can decompose the vector field into
simpler, component vector fields and examine whether in each case there appears a dominant behaviour associated with part of the field decomposition,
while the rest becomes subdominant in
some exact sense. Using this, we then built a system of integral
curves. The construction of these solutions eventually leads to a
formal asymptotic series expansion around the singularity and is done
term-by-term.

It is possible to describe the method of asymptotic splittings in an algorithmic way in three main steps as follows. Suppose we have a dynamical system of the sort $\dot{\mathbf{x}}=\mathbf{f}(\mathbf{x})$, with $(\cdot )\equiv d/dt$, or equivalently the vector field $\mathbf{f}$. We imagine that on approach to a finite time singularity, $\mathbf{f}$ decomposes into a dominant part $\mathbf{f}^{\,(0)}$ and another, subdominant part, as follows:
\begin{equation}
\mathbf{f}=\mathbf{f}^{\,(0)}+\mathbf{f}^{\,(1)}+\cdots +\mathbf{f}^{\,(k)}.
\label{dec1}
\end{equation}
This asymptotic decomposition is highly nonunique. Then, in step one of the method, for each particular decomposition we drop all terms after $\mathbf{f}^{\,(0)}$ and replace the exact equation  $\dot{\mathbf{x}}=\mathbf{f}(\mathbf{x})$  by the \emph{asymptotic equation}
\be
\dot{\mathbf{x}}=\mathbf{f}^{\,(0)}(\mathbf{x}).
\ee
Repeating this for all possible decompositions of $\mathbf{f}$, we end up with a variety of asymptotic dynamical systems to work with.

In Step two, we look for scale invariant solutions (called dominant balances) in each asymptotic system. Each such system may contain  many possible dominant balances resulting in various possible dominant behaviours near the singularity. Balancing each asymptotic system, requires a careful asymptotic analysis of the \emph{subdominant parts} in the  various  vector field decompositions.

In the last step, we check the overall consistency of the approximation scheme and build asymptotic solutions for each acceptable balance in a term-by-term iteration procedure that ends up with a formal series expansion representation of the solutions.

When the whole procedure is completed, we have finally constructed formal series developments of particular or general solutions of the original system of equations, valid in a local neighborhood of the finite time singularity. From such expansions we can:
 \begin{itemize}
   \item  deduce all possible dominant modes of approach of the field to the singularity
   \item  decide on the generality of the constructed solutions
   \item  determine the size and part of the space of initial data that led to such a solution.
 \end{itemize}

\subsection{Generic perturbations}\label{gp}
Formal series expansions can also be used in another way and address the problem of the possible \emph{genericity} of an exact solution, that is its placement when suitably perturbed in the solution space of the Einstein equations. This leads to the method of \emph{generic perturbations} expounded below. This method relies on certain function counting arguments and eventually on an initial value formulation of the theory under discussion.

We shall show below how this method works for the simple problem of finding out whether or not regular solutions in the form of formal asymptotic expansions are generic in the space of all solutions of general relativity in vacuum. The field equations $\textrm{Ric}=0$ in a Cauchy adapted frame split as follows:

\noindent\emph{\textsc{ADM} equations:}
\bq\label{adm}
\partial_t \g_{\GA\GB} &=& K_{\GA\GB},\nonumber\\
\partial_t K_{\GA\GB} &=&-2 P_{\GA\GB}  -\frac{1} {2}K K_{\GA\GB} +K^\GG_\GA K_{\GB\GG}.
\eq
\emph{Hamiltonian Constraint:}
\be
\mathcal{C}_0=P  +\frac{1}{4}K^2 -\frac{1}{4}K^{\GA\GB} K_{\GA\GB} =0.
\label{eq:hamiltonian}
\ee
\emph{Momentum Constraint:}
\be
\mathcal{C}_\a=\nabla_\b K^\GB_\GA - \nabla_\a K=0.
\label{eq:momentum}
\ee
Here $K=\textrm{tr}K_{\a\b}$, and $P_{\GA\GB}$ denotes the three-dimensional Ricci tensor associated with $\GG_{\GA\GB}$, and $P=\textrm{tr}P_{\a\b}$. The constraints show that the \emph{initial data} $(\g_{\a\b},K_{\a\b})$ cannot be chosen arbitrarily and must satisfy  the equations (\ref{eq:hamiltonian}) and (\ref{eq:momentum}) on each slice $\mathcal{M}_{t}$. The $\textsc{ADM}$ equations  describe the \emph{time development} $(\mathcal{V},g)$ of any initial data set $(\mathcal{M}_{t},\g_{\a\b},K_{\a\b})$ that satisfies the four constraint equations.

It is perhaps the simplest theorem in mathematical relativity that if we prescribe \emph{analytic} initial data $(\g_{\a\b},K_{\a\b})$ on some initial slice $\mathcal{M}_{0}$, then there exists a neighborhood of $\mathcal{M}_{0}$ in $\mathbb{R}\times\mathcal{M}$ such that the evolution equations  (\ref{adm}) have an analytic solution in this neighborhood consistent with these data. This analytic solution is the development of  the prescribed initial data on  $\mathcal{M}_{0}$ if and only if these initial data satisfy the constraints (\ref{eq:hamiltonian}) and (\ref{eq:momentum}) (see Ref. \cite{ycb} for a proof).

It is then not difficult to count the true degrees of freedom of the vacuum theory using the structure of the initial value problem. There are 12 initial data to be specified $(\g_{\a\b},K_{\a\b})$, but we have the four constraints  (\ref{eq:hamiltonian}) and (\ref{eq:momentum}) and also the freedom to perform 4 diffeomorphic changes, so the true number of free functions in the theory is $12-4-4=4$ in vacuum. This is the number that any solution of the vacuum Einstein equations must have if it is to qualify as a general solution.

Let us now assume a \emph{regular} formal series representation of the spatial metric of the form
\be
\GG_{\GA\GB}= \g^{(0)}_{\GA\GB} +\g^{(1)}_{\GA\GB}\;t + \g^{(2)}_{\GA\GB}\;t^2  + \cdots
\label{eq:3dimmetric}
\ee
where the $ \g^{(0)}_{\GA\GB} , \g^{(1)}_{\GA\GB} , \g^{(2)}_{\GA\GB} , \cdots$ are functions of the space coordinates. We shall be interested only in the part of the formal series shown, that is up to order two\footnote{Differentiation of such formal series with respect to either space or the time variables is defined term by term, whereas multiplication of two such expressions results when the various terms are multiplied and terms of same powers of $t$ are taken together.}. Thus \emph{before} substitution to the evolution and constraint  equations, the expression  (\ref{eq:3dimmetric}) contains  $18$ degrees of freedom. Note that setting $\g^{(0)}_{\GA\GB}=\d_{\a\b} $ and $\g^{(n)}_{\GA\GB}=0,n>0,$ we have Minkowski space included here as an exact solution of the equations, and so our perturbation analysis covers also that case.

The problem we are faced with in this subsection is what the initial number of 18 free functions becomes \emph{after} the imposition of the evolution and constraint equations, that is  how it finally compares with the 4 degrees of freedom that any general solution must possess as shown previously. Put it more precisely, given data $a_{\GA\GB} ,b_{\GA\GB} ,c_{\GA\GB}$, analytic functions of the space coordinates, such that the coefficients $\g^{(\mu)}_{\GA\GB}, \mu=0,1,2$ are prescribed,
\be\label{free}
 \g^{(0)}_{\GA\GB}= a_{\GA\GB}, \,\,  \g^{(1)}_{\GA\GB}=b_{\GA\GB}, \,\,   \g^{(2)}_{\GA\GB}= c_{\GA\GB},
\ee
how many of these data are truly independent when (\ref{eq:3dimmetric}) is taken to be a possible solution of the evolution equations  (\ref{adm}), together with the constraints  (\ref{eq:hamiltonian}) and (\ref{eq:momentum})?

For any tensor $X$, using the formal expansion  ($\ref{eq:3dimmetric}$), we can recursively calculate the coefficients in the regular  expansion
\be
X_{\GA\GB}= X^{(0)}_{\GA\GB} +X^{(1)}_{\GA\GB}\;t + X^{(2)}_{\GA\GB}\;t^2  + \cdots.
\label{x tensor}
\ee
For instance, using the first $\textsc{ADM}$ equation we find for the extrinsic curvature $K_{\a\b}$ that the coefficients  $K^{(n)}_{\GA\GB}$ of its regular expansion are given by the general recursive formal expression,
\be
\g^{(n)}_{\GA\GB}=\g^{(0)}_{\a\g}K^{(n)\,\g}_\b+\sum_{\mu+\nu=n}\g^{(\mu)}_{\GA\g}K^{(\nu)\,\g}_\b.
\ee
Using the field equations and calculating the various relations that appear at each order, we are led to the following ten relations between the data $a,b,c$: From the hamiltonian constraint (\ref{eq:hamiltonian}),  we get one relation, namely,
\be
c=\frac{1} {4}b^\GB_\GA b^\GA_\GB,
\label{eq:c}
\ee
from the momentum constraint (\ref{eq:momentum}), we obtain three more relations,
\be
\nabla_\b b^\GB_\GA  = \nabla_\a b,
\label{eq:gradb}
\ee
whereas from the space-space equations we get the following six relations,
\be
-P^\GB_\GA -\frac{1} {4}b^\GB_\GA b +\frac{1} {2}b^\GB_\GG b^\GG_\GA -c^\GB_\GA=0.
\label{eq:cab}
\ee
Hence, in total we find that the imposition of the field equations leads to 10 relations between the
  18 functions of the perturbation metric  (\ref{eq:3dimmetric}), that is we are left with 8 free functions. Taking into account the freedom we have in performing  4 diffeomorphism changes,  we finally conclude that there are in total 4 free functions in the solution  (\ref{eq:3dimmetric}). This means that  the regular solution (\ref{eq:3dimmetric}) corresponds to a general solution of the problem. Put it differently, `regularity is a generic feature of the Einstein equations in vacuum, under the assumption of analyticity'. We shall discuss  applications of this method in  later sections.

\subsection{Central projections}
A popular method to represent infinity is Penrose's conformal method (cf. Ref. \cite{pen}) wherein the overall structure of a physical spacetime is conformally changed so that its infinitely remote regions become the boundary of a new, unphysical spacetime. Infinity is then classified according to the behaviour of the various geodesics of the new spacetime near its boundary.

The method of asymptotic splittings expounded above in Section 2.3, offers an asymptotic representation of solutions to the field equations near infinity (that is where some component of the field diverges) but has two shortcomings:
\begin{enumerate}
  \item It does not give any information about the \emph{qualitative} behaviour of individual orbits
  \item It does not distinguish between the behaviour of those field components  tending to $+\infty$ from those ones  tending to $-\infty$.
\end{enumerate}
The qualitative method of central projections and compactifications due to Poincar\'e is usually presented\footnote{I thank Prof. J. Meiss for discussions of this point.}, cf. Refs. \cite{lef,per,dum,mei}, as suitable only for homogeneous vector fields and singularities at infinity, that is when we have a homogeneous vector  field diverging as $t\rightarrow\infty.$ However, this method is really suitable for any \emph{weight-homogeneous} vector field that blows up at a \emph{finite} time (or at infinity), and we may suitably adapt this ingenious method to study the finite-time  singularities that arise  in cosmology, and of course in any number of phase space dimensions (cf. Ref. \cite{sc13b} for details and applications).

For the case of the one-dimensional system $\dot{x}=f(x)$\footnote{for which we may assume  there is a finite time $t_s$ such that $x(t_s)=\pm\infty$}, the method of Poincar\'e is very briefly as follows\footnote{Note that in this one-dimensional case, the `field' $x$ has only one component.}. We imagine the one-dimensional phase space $\G=\{x:x\in\mathbb{R}\}$ of the system as sitting at the $(X,1)$ (cotangent) line of an $(X,Z)$ plane, and we let $x$ be any particular state of the system on this line (so that $x=X$ in this case). The Poincar\'e central projection $\textrm{pr}_c$ is a map from $\G$ to the upper semi-circle $\mathcal{S}_+^1$ sending each phase point to a point on the circle such that
\be
\textrm{pr}_c:\G\rightarrow\mathcal{S}_+^1:x\mapsto \theta ,\quad\textrm{with}\quad x=\cot\theta.
\ee
In this way, the two possible infinities of $x$ at the two ends of $\G$ are bijectively separated (note that this cannot happen with a stereographic map like Penrose's),
\bq
x=+\infty&\rightarrow&\theta=0\\
x=-\infty&\rightarrow&\theta=\pi,
\eq
the new phase space, the so-called \emph{Poincar\'e's circle}, $\mathcal{S}_+^1$ is now compact, $\theta\in [0,\pi]$, and the original system $\dot{x}=f(x)$ now reads,
\be
\dot\theta=-\sin^2\theta f(\cot\theta )\equiv g(\theta).
\ee
So past singularities are met as $\theta\downarrow 0$, while future ones are at $\theta\uparrow\pi$. Let us suppose for the sake of illustration that we are interested in  the asymptotic  properties of the field near a past singularity. In the next step of the Poincar\'e central projection method, we are after a singular \emph{asymptotic} system on the compactified phase space $\mathcal{S}_+^1$. According to the method of asymptotic splittings of Section 2.2,  the field $f$ is asymptotically decomposed into two parts, a dominant one $f^{(0)}$, and another subdominant, $f^{(\textrm{sub})}$, i.e., $f=f^{(0)}+f^{(\textrm{sub})}$, in a highly non unique way. We take any particular decomposition for which\footnote{As in the method of asymptotic splittings, we need to repeat the whole procedure below for all possible decompositions but, unlike that method, dominant balances play no role in  central projections.}
$
f^{(0)}=ax^m,
$
and therefore asymptotically as $\theta\downarrow 0$ we find that
$
g(\theta)=-a\theta^{2-m},
$
that is we arrive at the \emph{singular asymptotic system}
\be
\dot\theta=-a\theta^{2-m}.\ee
Notice that this system is only valid locally around the singularity at $\theta =0$, not everywhere on the compactified phase space $\mathcal{S}_+^1$. The last step is now to obtain a \emph{singularity-free} system valid locally around the singularity. For this purpose, one changes the time $t$ to a new one, $\tau$, given by $d/dt=\theta^{1-m}d/d\tau$, to arrive at the \emph{complete asymptotic system} defined on the Poincar\'e circle,
\be
\dot\theta=-a\theta,
\ee
and valid near $\theta=0.$  We conclude that the system has a past attractor when $a>0$, otherwise all orbits are repelled near the singularity.

To assess the merits of this method, we revisit the evolution  of the single-fluid FL models in general relativity with a linear equation of state $p=(\g-1)\rho$, studied  in Ref. \cite{we97}, pp. 58-60. This is governed by the equation
\be
\dot\Omega=-(3\g-2)(1-\Omega)\Omega,
\ee
and it is well known that for closed models, after a finite time interval, at the instant of maximum expansion, $\Omega$ blows up to $+\infty$, thus making the description of the whole evolution incomplete (cf. Ref. \cite{we97}, p. 60). To describe this, we choose the decomposition having $f^{(0)}=a\Omega^2$, where $a=3\g-2$ and $m=2$. The resulting  asymptotic system is $\dot\theta =-(3\g-2)$, and taking $\theta=0$ at $t=t_{\textrm{max}}$, we find that $\theta =-(3\g-2)(t-t_{\textrm{max}})$. The central projection gives $\Omega=\cot\theta=\cot(-(3\g-2)(t-t_{\textrm{max}}))$, near the point of maximum expansion.
\section{Extremophilic universes}
\subsection{Trapped surface formation}
A closed trapped surface is a 2-surface with spherical topology such that
both families of incoming and outgoing null geodesics orthogonal
to the surface converge. From the singularity theorems, cf. Ref. \cite{he73}, we know that a sufficient condition for the existence of an $\textsc{SF}_1$ type (collapse) singularity is the formation of closed trapped surface. The general problem of the conditions under which such a surface may form is probably one of the most important unsolved problems in general relativity (cf. Ref. \cite{christo} for recent progress on this very difficult issue).

In a cosmological setting, Ellis has shown in Ref. \cite{ellis} that a \textsc{RW}
space with scale factor $a(t)$ admits a past closed trapped
surface if the following condition is satisfied: $
\dot{a}(t)>\left|{f'(r)}/{f(r)}\right|,$ with $f(r)=\sin
r,r,\sinh r$ for $k=1,0,-1$ respectively. In terms of the Bel-Robinson energy, this condition reads
\be \label{d}
|D|>\frac{\sqrt{3}}{a^{2}(t)f^{2}(r)} ,
\ee
and we conclude that $\textsc{SF}_1$ singularities (as predicted by the existence of a trapped
surface) are characterized by a divergent Bel-Robinson energy (sufficient condition). The standard dust and radiation big bang singularities are examples of a $(\textsc{SF}_1,\textsc{BR}_2,\textsc{H}_2)$ type, cf. Refs. \cite{ck07,split}.

Closed trapped surfaces also exist (as shown in Ref. \cite{cot06}) in graduated inflationary universes, cf. Ref. \cite{ba}, in the flat models with a scalar field of Ref. \cite{fo}, as well as in  models with sudden singularities.

\subsection{Sudden singularities}
A sudden singularity will be said to arise everywhere at comoving proper
time $t_{s}$ in a Friedmann universe expanding with scale factor $a(t)$ if
\begin{equation}
\lim_{t\rightarrow t_{s}}a(t)=a_{s}\neq 0,\qquad \lim_{t\rightarrow t_{s}}%
\dot{a}(t)=\dot{a}_{s}<\infty ,\qquad \lim_{t\rightarrow t_{s}}\ddot{a}%
(t)=\infty,
\end{equation}
for some $t_{s}>0$. Barrow in Ref. \cite{ba04} discovered a  new class of pressure-driven
singularities that keep the scale factor, $a$, expansion rate, $\dot{a}/a$,
and the density, $\rho $, finite while the pressure, $p$, blows up at a
finite time despite the energy conditions $\rho >0$ and $\rho +3p>0$
holding. Their status as stable solutions of the classical Einstein
equations in the presence of small scalar, vector and tensor perturbations
has been studied in a gauge covariant formalism in Ref. \cite{lip}, and they have
also been found to be stable against quantum particle production processes in Ref.
\cite{fab}.

The original solutions of Barrow in the references cited above constitute a 1-parameter family of solutions of the field equations. In Ref. \cite{sc10a} we have constructed them as asymptotic series solutions. This analysis was helpful to locate the precise position of the arbitrary constant in the series, while we also gave asymptotic forms of the diverging pressure near the sudden singularity. Is there a general solution of the Friedman equations with a sudden singularity? Is there a general solution of the full Einstein equations with  a sudden singularity? In Ref. \cite{sc10b} we have given a positive answer to the second question by constructing an asymptotic series solution to the full gravitational equations near a sudden singularity. In this solution the density, expansion rate, and metric
remain finite, the mean scale factor
and the shear approach constant values, the chaotically anisotropic degrees
of freedom which dominate at an initial curvature singularity are frozen
out, and anisotropic velocities grind to a halt because of the divergent
pressure and inertia. The situation is far simpler than for a initial vacuum
or $p<\rho $ dominated fluid singularity. Comparable simplicity is achieved
in the non-singular late-time approach to a quasi-isotropic de Sitter
universe in the presence of a cosmological constant or a $p=-\rho $ fluid,
shown by Starobinsky in Ref. \cite{nohair}, and in the regular states of higher order gravity (cf. Ref. \cite{sc12d} and also Section 4.3  below). As to the first question, a positive answer is in sight, cf. Ref. \cite{sc13c}, where other results of a deeper interest are included.
\subsection{Dark energy}
A representative family of models with an $\textsc{SF}_3$ type (big rip) singularity is the class of universes with diverging pressure studied in Ref. \cite{ck07} (cf. Thm. 3.1 in that reference). Suppose we have  an
isotropic universe, flat or curved, filled with a fluid with  equation of state
$p=w\rho$ such that $w<-1$ and $|p|\rightarrow\infty$ at the finite time $t_{s}$, the position of the singularity. From the continuity equation it follows that $\rho\propto a^{-3(w+1)},$ and so if $w<-1$ and
$p$ blows up at $t_{s}$, $a$ also blows up at $t_{s}$. This is a big rip singularity. Further, we find that $|D|$ and $|E|$ are divergent signaling a $\textsc{BR}_2$ type. Since the curvature term in the Friedman equation is asymptotically negligible in this model, and the density blows up, $H$ is also diverging, meaning an $\textsc{H}_2$ type singularity. Collectively, this family of models has a big rip singularity of the type $(\textsc{SF}_3,\textsc{BR}_2,\textsc{H}_2)$. A physical example of a universe belonging to this family of models is the flat, phantom dark energy model of Ref. \cite{gonzales}.

\subsection{Interacting fluids}
In recent years there have been an increasing number of works devoted to analyzing diverse problems in situations involving  more than one cosmological fluids that show \textit{a mutual interaction} and the associated exchange of energy. In particular, it is important to  understand the nature of finite-time singularities and more generally the possible asymptotics that may be admitted in cosmological models with interacting fluids. Such an understanding will complement current physical studies of such models which focus on other issues and may also provide a demarcation of the range of dynamical possibilities of these models.
\subsubsection{Interacting fluids: The flat case}
We have made the first steps in this direction in Ref. \cite{sc12a} by providing  an analysis of the singular phenomena that emerge when we consider two interacting perfect fluids with equations of state $p_1=(\G
-1)\r_1, p_2=(\g -1)\r_2$ in a \emph{flat} FRW universe. The field equations for such a universe eventually reduce to the following cubic system for $x=H$:
\be\label{cubic}
\dot x=y,\quad
\dot y=-Axy-Bx^3.
\ee
Here $\a,\b$ are signed constants signifying the mutual transfer  of energy, and $A=\a+\b+3\g+3\G$,
$B=3(\a\G+\b\g+3\G\g )/2.$
We have examined what happens when we take this system asymptotically to a finite-time singularity and found a number of regimes described by different asymptotic solutions - seven different behaviours in all. These lead to either collapse or  big rip  singularities, with the possible exception of the case of phantom matter wherein only asymptotics leading to a big bang type behaviour are possible. It is interesting that a deeper study of the cubic system (\ref{cubic}), cf. Ref. \cite{sc13b}, shows that the qualitative structure of the phase space has an elliptic domain for finite orbits, whereas the equilibria at infinity can be very complex and include parabolic, elliptic and hyperbolic sectors in the first quadrant of the $A,B$ space.

\subsubsection{The asymptotic influence of  curvature}
The richness of the singular  behaviours summarized above for the flat case, makes the dynamics of  cosmologies with two interacting fluids especially interesting on approach to their singularities, so that a number of closely related problems may be examined. The first is  precisely how the inclusion of curvature alters the behaviours found in the flat case and whether new and distinct forms are possible. In this case, the system of equations becomes \emph{three-dimensional} rather than the two-dimensional one we had in the flat case above. One interesting aspect of the whole analysis is  the fact that there are asymptotic solutions both when the curvature term is included in the dominant part of the vector field  on approach to the singularity, \emph{and also} when the curvature term exists only in the subdominant part asymptotically. In either case, there are two new qualitative features that appear when curvature influences the dynamics besides those already present in the flat case (which continue to exist in the case with curvature), cf. Ref. \cite{kittou2}:
\begin{enumerate}
\item The existence of horizon breaking solutions asymptotically at early times
\item The appearance of sudden singularities.
\end{enumerate}
\subsubsection{More general interactions}
There are various cases of great physical (and mathematical) interest to study when we allow for more general interactions between the two fluids, for example  when the original system of Friedman equations assumes the form,
\bq\label{sys}
3H^2&=&\r_1+\r_2\nonumber\\
\dot{\r}_1+3H\G\r_{1}&=&-\beta H^m\r_{1}^{\l} +\a H^n\r_2^{\,\m}\\
\dot{\r}_2+3H\g\r_2&=&\beta H^m\r_1^\l -\a H^n\r_2^{\,\m}.\nonumber
\eq
This system is currently under close scrutiny, cf. Ref. \cite{kittou3}. An important new qualitative feature that appears when we allow for such generalized interacting fluids is the appearance of the fluid parameters $\G,\g$ in the dominant forms of the asymptotic, scale-invariant solutions. Hence, we may be able to control the singularity types not only through  changing the type of fluid or turning on and off the asymptotic influence of the curvature, but also through the managing of the way the two fluids interact with one another.

\section{Higher order gravity}
Studying infinity in higher order gravity and scalar-tensor theory presents before us two major challenges,  \emph{late-time stability}-that is deciding the fate of these universes in the distant future, and \emph{early-time stability}-examining the past evolution at early times towards a possible initial singularity. There is also a related bundle of ideas, that of studying the \emph{genericity} of exact solutions of the theory. In this Section, we review recent work on these topics and also present some results of related ongoing research.
\subsection{Radiation universes}
Our starting point is the metric (\ref{rwmetrics}) filled  with a radiation fluid with energy-momentum tensor
$T_{\mu\nu}=(p+\rho)u_\mu u_\nu +pg_{\mu\nu}$,
where the fluid velocity 4-vector is $u^\mu=\delta^\mu_0$, and we take the equation of state to be $p=\rho/3$. This is combined with the higher order action
\be\mathcal{S}=\frac{1}{2}\int_{\mathcal{M}^4}\mathcal{L}_{\textrm{total}}d\mu_{g},
\ee
where $\mathcal{L}_{\textrm{total}}$ is the lagrangian density of the general
quadratic gravity theory given in the form $\mathcal{L}_{\textrm{total}}=\mathcal{L}(R)+\mathcal{L}_{\textrm{matter}}$, with
\be
\mathcal{L}(R)=R + \b R^2 + \g \textrm{Ric}^2 + \d \textrm{Riem}^2 ,
\label{eq:lagra}
\ee
where $\b,\g,\d$ are constants. This  leads to the cosmological equations
\be
\frac{k+\dot{a}^2}{a^2}+\xi\left[2\: \frac{\dddot{a}\:\dot{a}}{a^2} + 2\:\frac{\ddot{a}\dot{a}^2}{a^3}-\frac{\ddot{a}^2}{a^2} - 3\:
\frac{\dot{a}^4}{a^4} -2k\frac{\dot{a}^2}{a^4} + \frac{k^2}{a^4}\right] = \frac{\zeta^2}{a^4} ,
\label{eq:beq}
\ee
where  $\xi=2(3\b+\g)$ and $\zeta$ is a constant defined by the constraint
\be
\frac{\rho}{3}=\frac{\z^2}{a^4},\quad (\textrm{from}\,\,\nabla_{\mu}T^{\mu 0}=0).
\ee
Setting $x=a$, $y=\dot{a}$ and $z=\ddot{a}$, Eq. (\ref{eq:beq}) can be written as an autonomous dynamical
system of the form
\be\label{basic dynamical system}
\mathbf{\dot{x}}=\mathbf{f}_{\,k,\textsc{RAD}}(\mathbf{x}),\quad \mathbf{x}=(x,y,z),
\ee
that is we have the dynamical system
\begin{eqnarray}
\label{eq:ds}
\dot{x} &=& y,\:\:\:\:\:\nonumber\\ \dot{y} &=& z,\:\:\:\:\:\\
\dot{z}& =& \frac{\z^2-k^2\xi}{2\xi x^2y} + \frac{3y^3}{2x^2} + \frac{z^2}{2y} -\frac{yz}{x} - \frac{y}{2\xi}
-\frac{k}{2\xi y} + \frac{ky}{x^2},\nonumber
\end{eqnarray}
 equivalent to the \emph{curvature-radiation} vector field $\mathbf{f}_{\,k,\textsc{RAD}}:\mathbb{R}^3\rightarrow\mathbb{R}^3:(x,y,z)\mapsto\mathbf{f}_{\,k,\textsc{RAD}}(x,y,z)$ with
\be\label{vf}
\mathbf{f}_{\,k,\textsc{RAD}}(x,y,z)=\left( x,y,\frac{\z^2-k^2\xi}{2\xi x^2y} + \frac{3y^3}{2x^2} + \frac{z^2}{2y} -\frac{yz}{x} - \frac{y}{2\xi}
-\frac{k}{2\xi y} + \frac{ky}{x^2}\right).
\ee
The curvature-radiation field $\mathbf{f}_{\,k,\textsc{RAD}}$  combines the effects of curvature and radiation and describes completely the dynamical evolution of any radiation-filled FRW universe in higher order gravity.

We recall that when $k=0$, that is when we have a flat, radiation-filled  FRW model, the vector field $\mathbf{f}_{\,0,\textsc{RAD}}$ has two admissible asymptotic solutions near the initial singularity, as shown in Ref. \cite{K1}: In the first family, all flat, radiation solutions are dominated (or attracted) at early times by the form $a(t)\sim t^{1/2}$, thus proving the stability of this solution in the flat case. There is a second possible asymptotic form near the singularity in the flat case, $a(t)\sim t$, but this contains only two arbitrary constants and hence it corresponds to a \emph{particular} solution of the theory (cf. \cite{K1}).

When $k\neq 0$,  the field $\mathbf{f}_{\,k,\textsc{RAD}}$ has more terms (those that contain $k$ in (\ref{vf})) than in the flat case. (To count exactly how many terms there are, see Ref. \cite{kolionis1}.) Proceeding with the method of asymptotic splittings, in  Ref. \cite{kolionis1} we were able to show that asymptotically the effects of curvature and radiation on the global evolution become negligible and the initial state of these universes is effectively described by the flat, vacuum $t^{1/2}$ solution of these theories (with the possible exception of the solutions in the conformally invariant Bach-Weyl theory - this case apparently needs a separate analysis). In particular, all curved radiation solutions tend to this solution asymptotically as we approach the initial state in these theories. Using various asymptotic and geometric arguments, we were able to built a solution of the field equations in the form of a Puiseux formal series expansion compatible with all other constraints, dominated asymptotically by the $t^{1/2}$ solution and having the correct number of arbitrary constants that makes it a general solution. In this way, we can conclude that this exact solution is an attractor of all homogeneous and isotropic radiation solutions of the theory thus proving stability against such `perturbations'.
\subsection{Vacuum models}
The impressive restrictions placed by the higher order field equations on the structure of the possible initial cosmological states of the theory imply that the initial state of the radiation universes studied above effectively resembles that of a vacuum, flat  model,  since  the only possible mode of approach to the singularity is one  in which both the curvature \emph{and} radiation modes enter asymptotically only in the subdominant part of the vector field. In fact, an investigation of the asymptotic stability of the flat, vacuum initial state in these theories with respect curved FRW perturbations is currently under scrutiny, cf. \cite{kolionis2}. For negatively curved universes, we find that asymptotically they are all dominated by a stable Milne-type cosmological solution, and there are no other possible dominant features on approach to the singularity in these models. These results also allow us to conclude a global singularity theorem valid conformally with the initial state of the Einstein-scalar field system characterized by the scalar field blowing up like $\ln t$, and the Hubble parameter diverging as $1/t$.
\subsection{Generic regular states}
If both states prove to be stable, then whether or not the radiation or the vacuum initial states are in addition \emph{generic}  in the space of solutions of the higher order gravity equations is currently an open problem. In fact, only very recently it was shown in Ref. \cite{trachilis1} that the higher order gravity field equations in vacuum admit a unique solution in the form of a \emph{regular} formal power series expansion which contains the right number of free functions to qualify as a general solution of the system. This required using the method of generic perturbations discussed above in Section 2.3, including a careful function counting technique, and it was also necessary to develop a formulation of the theory as a system of evolution equations with constraints.

In fact, there is a similar, but more complicated, ADM-like structure with constraints of the theory derived from the quadratic  lagrangian  $f(R)=R+\b R^2$ and satisfied by any initial data set of the form $(\mathcal{M}_{t},\g_{\a\b},K_{\a\b},D_{\a\b},W)$  consisting of the spatial metric $\g_{\a\b}$, the extrinsic curvature $K_{\a\b}$, the acceleration tensor $D_{\a\b}$ and the jerk tensor $W_{\a\b}$. Further, as shown in Ref. \cite{trachilis1}, the whole system of evolution equations and constraints is of the Cauchy-Kovalevski type, and if we prescribe initial data analytic on some initial slice $\mathcal{M}_{0}$, then there exists a neighborhood of $\mathcal{M}_{0}$ in $\mathbb{R}\times\mathcal{M}$ such that the evolution equations have an analytic solution in this neighborhood consistent with these data. This analytic solution is the development of  the prescribed initial data on  $\mathcal{M}_{0}$ if and only if these initial data satisfy the constraints. In this way, we are driven to count the true degrees of freedom of the higher order gravity theory we consider:  There are  19 evolution equations subjected to 4 constraints and we also have the freedom to perform 4 diffeomorphic changes. Thus in total we have in vacuum $19-4-4=11$ degrees of freedom. This in turn implies that any solution with 11 free functions has the same degree of generality with a general solution of the theory.

Following the method of generic perturbations of Section \ref{gp} for the more complicated higher order equations, we look for a 30 degrees of freedom regular formal series representation of the spatial metric of the form
\be
\GG_{\GA\GB}= \g^{(0)}_{\GA\GB} +\g^{(1)}_{\GA\GB}\;t + \g^{(2)}_{\GA\GB}\;t^2 + \g^{(3)}_{\GA\GB}\;t^3 + \g^{(4)}_{\GA\GB}\;t^4 + \cdots .
\label{eq:3dimmetricHOG}
\ee
We ask: What will the initial number of thirty free functions become after the imposition of the higher order evolution and constraint equations, that is  how it finally compares with the 11 degrees of freedom that any general solution must possess? Put it more precisely, given data $a_{\GA\GB} ,b_{\GA\GB} ,c_{\GA\GB} ,d_{\GA\GB} ,e_{\GA\GB}$, analytic functions of the space coordinates, such that the coefficients $\g^{(\mu)}_{\GA\GB}, \mu=0,\cdots 4,$ are prescribed,
\be\label{free}
 \g^{(0)}_{\GA\GB}= a_{\GA\GB}, \,\,  \g^{(1)}_{\GA\GB}=b_{\GA\GB}, \,\,   \g^{(2)}_{\GA\GB}= c_{\GA\GB}, \, \,   \g^{(3)}_{\GA\GB}= d_{\GA\GB}, \,\,
     \g^{(4)}_{\GA\GB}= e_{\GA\GB},
\ee
how many of these data are truly independent when (\ref{eq:3dimmetricHOG}) is taken to be a possible solution of the evolution equations   together with the constraints? The theorem we proved in Ref. \cite{trachilis1} concerning this problem is the following.
\begin{theorem}
Let $a_{\a\b}$ be a smooth Riemannian metric , $b_{\a\b},c_{\a\b},d_{\a\b}$ and $e_{\a\b}$  be symmetric smooth tensor fields which are traceless  with respect to the metric $a_{\a\b}$, i.e., they satisfy $b=c=d=e=0$.  Then there exists a formal power series expansion solution of the vacuum higher order gravity equations of the form (\ref{eq:3dimmetricHOG}) such that:
\begin{enumerate}
\item It is unique
\item The coefficients $\g^{(n)\,\GA\GB}$ are all smooth
\item It holds that $\g^{(0)}_{\GA\GB}=a_{\a\b}$ and $\g^{(1)}_{\GA\GB}=b_{\a\b},$ $\g^{(2)}_{\GA\GB}=c_{\a\b}$, $\g^{(3)}_{\GA\GB}=d_{\a\b}$ and  $\g^{(4)}_{\GA\GB}=e_{\a\b}$.

\end{enumerate}
\end{theorem}
This result shows that if we start with an analytic initial data set in which the metric has the asymptotic form (\ref{eq:3dimmetricHOG}) and evolve, then we can build an asymptotic development  in the form of a formal series expansion which satisfies the evolution and constraint equations and has the same number of free functions as those of a general solution of the theory. In other words, we have shown that regularity is a generic feature of the $R+\b R^2$ theory under the assumption of analyticity.

\section{Brane cosmologies}
In  Refs. \cite{kl0a,kl0b,kl0c,kl1,kl2}, we have elaborated on the dynamics of the so-called self-tuning mechanism introduced in  Refs. \cite{nima,silver}. This is an approach to the cosmological constant problem using ideas imported  from brane theory, and aiming to examine in a model-independent way the possibility of \emph{avoiding} singularities in the bulk at a finite distance from the brane position. The general problem depends on the five dimensional Einstein equations
\be
G_{AB}=\kappa^{2}_{5}T_{AB},
\ee
where $\kappa^{2}_{5}=M_{5}^{-3}$
and $M_{5}$ is the five dimensional Planck mass, which for the geometric setup of the metric (1.4) described in the Introduction, can  be eventually  written as
\bq
\label{orig_system1}
\dfrac{a''}{a}&=&-A\lambda\phi'^{2}-\dfrac{2}{3}A(1+2\gamma)\rho_{2},\\
\label{orig_system2}
\dfrac{a'^{2}}{a^{2}}&=&\dfrac{\lambda
A}{3}\phi'^{2}+ \dfrac{2A}{3} \rho_{2}+\dfrac{kH^{2}}{a^{2}}, \eq where
$A=\kappa_{5}^{2}/4$, $k=0,\pm 1$ (and the prime ($'$) denotes
differentiation with respect to $Y$). Eqs. (\ref{orig_system1}) and (\ref{orig_system2}) are not independent, since
Eq. (\ref{orig_system1}) was derived after substitution of Eq. (\ref{orig_system2}) in
the field equation $G_{\a\a}=\kappa_{5}^{2}T_{\a\a}=4 A T_{\a\a}$, $\a=1,2,3,4$:
\be
\frac{a''}{a}+\frac{{a'}^{2}}{a^{2}}-\frac{k H^{2}}{a^{2}}=-\frac{2A}{3}\lambda{\phi'}^{2}-
\frac{4A}{3}\gamma\rho_{2}.
\ee
In general, there is an exchange of energy between the two matter components depending on the values and signs of the two constants $\nu,\sigma$ appearing below, such that the total energy is conserved,  and we have the following two equations,
\bq \label{orig_system3}
\lambda\phi'\phi''+4\lambda\dfrac{a'}{a}{\phi'}^{2}&=&-\frac{\lambda\nu}{2}
\frac{a'}{a}{\phi'}^{2}+\sigma\rho_{2}\frac{a'}{a},\\
\label{orig_system4}
\rho_{2}'+4(\gamma+1)\dfrac{a'}{a}\rho_{2}&=&\frac{\lambda\nu}{2}
\frac{a'}{a}{\phi'}^{2}-\sigma\rho_{2}\frac{a'}{a}.
\eq
In our analysis in Refs. \cite{kl0a,kl0b,kl0c,kl1,kl2}, we use the independent Eqs. (\ref{orig_system1}), (\ref{orig_system3}) and (\ref{orig_system4})
to determine the unknown variables $a$, $a'$, $\phi'$ and $\rho_{2}$,
while Eq. (\ref{orig_system2}) plays the role of a constraint equation.

For either  flat or  curved brane configurations, there are three main  cases to be analyzed according to whether the two matter components are either independent, or coexist without exchange of energy, or finally, interact with each other. In general the singularities, when they exist, are either of an $\textsc{SF}_1$-type (collapse), or of type $\textsc{SF}_3$ (big rip). In particular, there are no generic sudden singularities in all such models. For a pure scalar field bulk, cf. Ref. \cite{kl0a,kl0b,kl1}, there are only $\textsc{SF}_1$-type singularities and the dynamical behaviour strongly depends on  whether the brane is flat or not. The found singularity in the flat case is of type $\textsc{BR}_2$, cf. Refs. \cite{kl0a,kl0b,kl1} (see also Ref. \cite{nima}), and as the model approaches it, all the vacuum energy decays (since $\phi'\rightarrow\infty$,
as $Y\rightarrow Y_{s}$). For a curved brane, the model avoids the singularity which is now located at an infinite distance.

The existence of a
perfect fluid instead of (or coexisting but not interacting with) a scalar field in the bulk enhances (as expected!) the dynamical possibilities of
brane evolution in the fluid bulk, cf. Refs. \cite{kl0c,kl1}. Such possibilities stem from the
different possible asymptotics  of the fluid density and the
derivative of the warp factor with respect to the extra dimension. In contrast to the bulk scalar field case where all curved brane solutions were regular, now we found also \emph{singular}  solutions.
The results depend crucially on the values of the parameter $\gamma$:  (i) For $\gamma>-1/2$, the flat brane
solution develops a collapse singularity at finite distance,
that disappears in the curved case. (ii) For $\gamma<-1$, the
singularity cannot be avoided and it changes to a big rip for
a flat brane. (iii) For $-1<\gamma\le -1/2$, the surprising result
is found that while the curved brane solution is singular, the flat
brane is not, opening the possibility for a revival of the
self-tuning proposal. Thus, the self-tuning mechanism appears to be a property of a general
(non-singular) flat brane solution, that depends on \emph{two} arbitrary constants in the region $-1<\gamma<-1/2$
(three for the general solution in an exceptional case with sudden behavior when $\gamma=-1/2$).

Lastly, when the matter components are an interacting  mixture of
an analog of perfect fluid and a massless scalar field, we have the most complicated case to be checked for any new behaviours. In a particular range of the interaction parameters, we found in Ref. \cite{kl2} flat brane general solutions avoiding the singularity at finite distance
from the brane, in the same region of the equation of state constant parameter
$\gamma=P/\rho$ that we found previously in the absence of the bulk scalar field
$(-1<\gamma<-1/2)$.

It would be an interesting deeper problem to analyze the qualitative behaviour of the orbits through central projections in various of these brane models. This could lead to a better understanding of how the found exact behaviours are realized.

\section*{Acknowledgements}
I thank my collaborators Ignatios Antoniadis, John Barrow, Yvonne Choquet-Bruhat, Georgia Kittou, Ifigeneia Klaoudatou, Giorgos Kolionis, John Miritzis, Dimitrios Trachilis and  Antonios Tsokaros for their friendly cooperation, fertile exchange of ideas and keen interest in the adventurous explorations of the emerging universes of ideas summarized here.

\end{document}